\documentclass{IEEEcsmag}

\usepackage[colorlinks,urlcolor=blue,linkcolor=blue,citecolor=blue]{hyperref}

\usepackage{graphicx} 
\graphicspath{{Figures/}} 

\usepackage{subcaption}

\usepackage{upmath}
\usepackage{cite}

\jvol{XX}
\jnum{XX}
\paper{8}
\jmonth{May/June}
\jname{IT Professional}
\pubyear{2021}

\begin{document}

\sptitle{}

\title{Reflections on the Use of Dashboards in the Covid-19 Pandemic}

\author{A. Arleo}
\affil{Eindhoven University of Technology, Eindhoven, 5600 MB, The Netherlands\\and Vienna University of Technology, Vienna, A-1040, Austria}

\author{R. Borgo}
\affil{King's College, London, WC2B 4BG, United Kingdom}

\author{J{\"o}rn~Kohlhammer}
\affil{Fraunhofer IGD and TU Darmstadt, Germany}

\author{Roy A Ruddle.}
\affil{University of Leeds, Leeds, LS2 9JT, United Kingdom}

\author{H. Scharlach}
\affil{Public Health Agency of Lower Saxony, Hannover, 30449, Germany}

\author{Xiaoru Yuan}
\affil{Peking University, Peking, 100191, People's Republic of China}

\markboth{}{}

\begin{abstract}
\looseness-1 Dashboards have arguably been the most used visualizations during the COVID-19 pandemic. They were used to communicate its evolution to national governments for disaster mitigation, to the public domain to inform about its status, and to epidemiologists to comprehend and predict the evolution of the disease. Each design had to be tailored for different tasks and to varying audiences - in many cases set up in a very short time due to the urgent need. In this paper, we collect notable examples of dashboards and reflect on their use and design during the pandemic from a user-oriented perspective:
we interview a group of researchers with varying visualization expertise who actively used dashboards during the pandemic as part of their daily workflow. We discuss our findings and compile a list of lessons learned to support future visualization researchers and dashboard designers.

\end{abstract}

\maketitle

\chapteri{``F}lattening the curve." During the Covid-19 pandemic, this expression became representative of every effort oriented to contain the spread of the virus. Whenever new restrictive measures were introduced by the government, they were usually motivated and justified as necessary sacrifices to finally flatten the line on the charts that were every evening on the news. The fact that such widespread expression came from a \textit{visual} representation of the pandemic evolution tells a lot about the role that visualization had in informing the public and supporting experts throughout these difficult times. 

Dashboards combine different interactive views to provide the users an easy to use, compact yet profound exploration tool for complex datasets~\cite{few2006information, kitchin2015knowing, bach2023dashboard}. They are designed to enable access to large amounts of information in a time-efficient way and have been used in countless application domains~\cite{bach2023dashboard}. These factors contributed to their success as the go-to tool for exploring pandemic data, 
and close to every country in the world created its own 
to track the pandemic evolution, communicate with the public, and support disaster mitigation~\cite{zhang2023dashboards}.  

Previous research on dashboard design focused on collecting  best practices from designers and creators~\cite{bach2023dashboard,zhang2023dashboards}. 
In this paper, we explore a different angle, analyzing the COVID-19 dashboard \textit{phenomenon} from a ``field-user" point of view. We discuss successes and pitfalls, and interview experts who used dashboards on the ``front line", with most of them approaching visual analysis of data, beyond simple tables and charts, for the first time. This gives us a unique perspective, that complements and expands over existing literature. 
We complement these findings by discussing the developers' view in this context.
The main contribution of our paper consists of a discussion about the insights we obtained during our interviews with domain experts who share their experience with dashboards during the pandemic. We reflect on a set of recommendations aimed at supporting upcoming design and development of dashboards in the context of prevention of (and preparation for) future pandemic events.  The paper has been conceived during the Dagstuhl Seminar 24091 about ``Reflections on Pandemic Visualization".






\section{Dashboards in Visualization}

The term ``dashboard" was originally used to define any vehicle's driver instrument panel. The principle behind a dashboard is providing a comprehensive view of several indicators in a succinct space. This type of information visualization later found its way into different domains, including finance, healthcare, and many others~\cite{few2006information, sarikaya2018we}. Wexler et al. defines them as ''\textit{a visual display of data used to monitor conditions and/or
facilitate understanding}"~\cite{wexler2017big}. 

The concept of dashboards evolved from single page/view reporting screens to visual analytics powerhouses, including interactivity and different levels of detail. Sarikaya et al.~\cite{sarikaya2018we} survey 83 dashboards, categorizing them by purpose, audience, visual features, and data semantics. They also discuss lessons from dashboards ``in use". However, these were derived primarily from literature review.

Bach et al.~\cite{bach2023dashboard} survey 144 dashboards and categorize design patterns into eight categories, which can be broadly divided in \textit{content}, concerning data abstraction, and \textit{composition}, concerning information arrangement, interactions and use of color.

Zhang et al.~\cite{zhang2023dashboards} surveyed dashboard design practices during COVID-19 ``behind the scenes" with their creators. Similar to our intended goal, but yet with a different point of view, this paper reports on the visualization creation process during COVID-19 by interviewing the dashboard creators. The result is a collection of findings and insights that provide an original yet insightful panorama of the socio-technical context and human processes behind visualization development, maintenance, and termination.

\begin{figure}[t!]
    \centering
    \includegraphics[width=0.7\linewidth, trim={3cm 0 3cm 0}, clip]{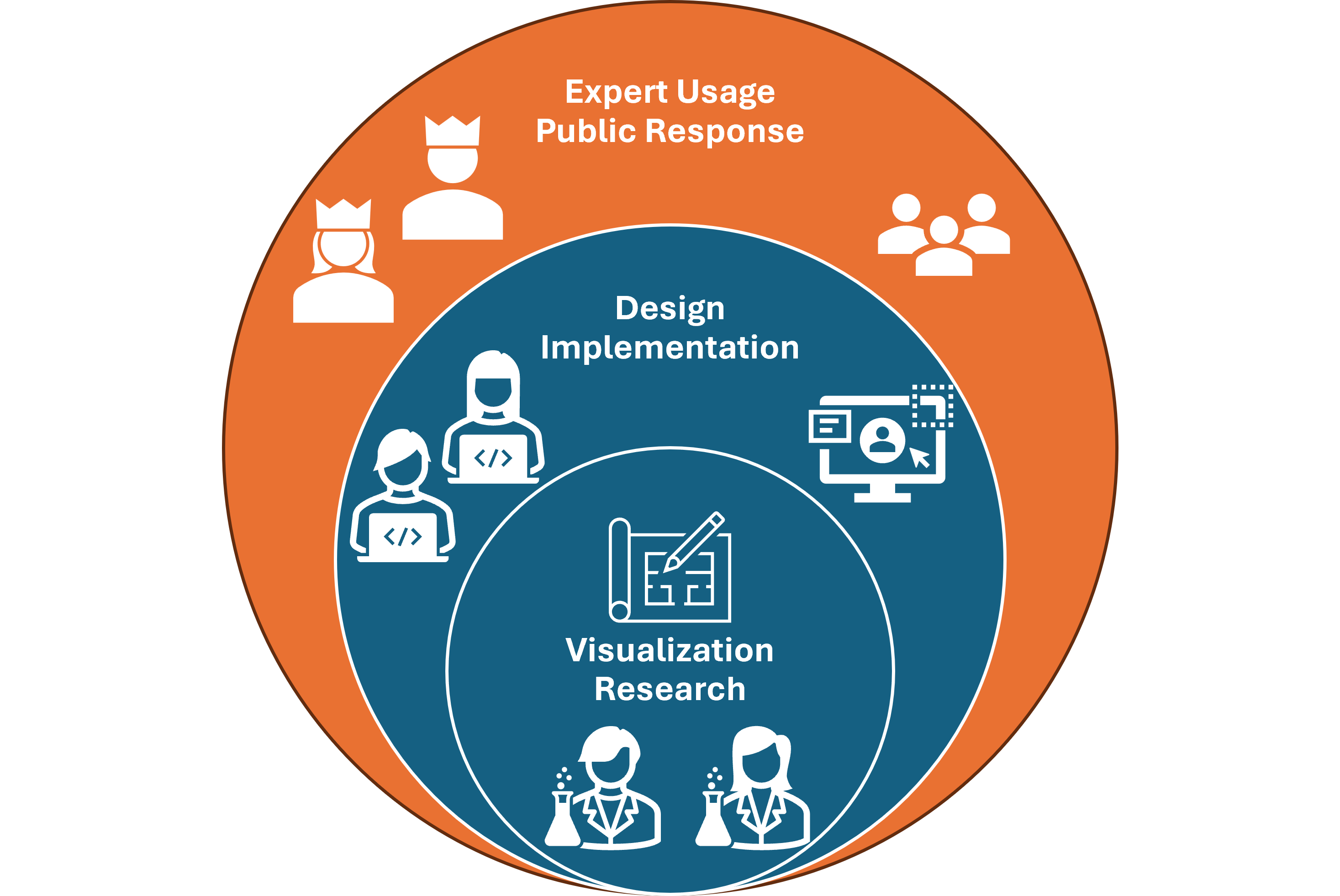}
    \caption{Framing our paper contribution (orange) within the visualization design process. }
    \label{fig:framework}
\end{figure}

We frame our contribution in the state-of-the-art as in Figure~\ref{fig:framework}. The paper by Bach et al.~\cite{bach2023dashboard} acts as a theoretical foundation for dashboard design, while Zhang et al.~\cite{zhang2023dashboards} provide a perspective on the actual creation process from the creator themselves. Our contribution provides anecdotal evidence about the usage of the dashboards by experts and public response.


\section{Dashboard Design during COVID-19: our Developer Perspective}

In our paper, we differentiate from the existing literature as we focus and prioritize the user point-of-view, and we base our reflections from real-world experiences described in the own words of the protagonists. 
We are particularly interested on the experience with COVID-19 dashboards of professionals with no specific visualization background. Nonetheless, there are also a few aspects that were important in the specific context of COVID-19 dashboard development concerning the visualization design process - and for the sake of completeness  we briefly discuss them in the following. 

\textit{Time to develop.} The time to create the visualizations based on previous experience was relatively low (as low as 1 person month). Medical experts and epidemiologists were a scarce resource (depending on the phase of the pandemic), making a structured user-centered design process or a dedicated analysis of data, users, and tasks ~\cite{miksch2014triangle} highly demanding.

\textit{Audience.} In the specific case of dashboard development for the general public, the selection of idioms proved a more difficult task than usual. When the audience base is limited to a specific circle of experts or a single category, it is possible to predict their visualization literacy - and it is also easier to conduct iterative improvements based on their feedback. In the case of approaching the general public, a varied and vast audience, this process becomes orders of magnitude more difficult. As an added challenge, ambiguous encodings might lead to misinterpretation of the data, leading to disinformation on a national-scale level.

\textit{Developers arrangement.} There was a wide range of development team setups. The list of dashboards that the Dagstuhl seminar examined is biased towards those that included dedicated visualization experts in their development teams. However, many Covid-19 dashboards were developed without an explicit reference to visualization experts.

\section{Dashboards in the Field}\label{se:dashboard_review}

A huge number of dashboards were developed during the pandemic. They differed in the technology applied, the data displayed and the user groups in focus. We identified four different types of dashboard/report examples that we used as a basis for the interviews.

\begin{figure*}[t!]
    \begin{subfigure}{0.5\textwidth}
        \includegraphics[width=0.9\linewidth]{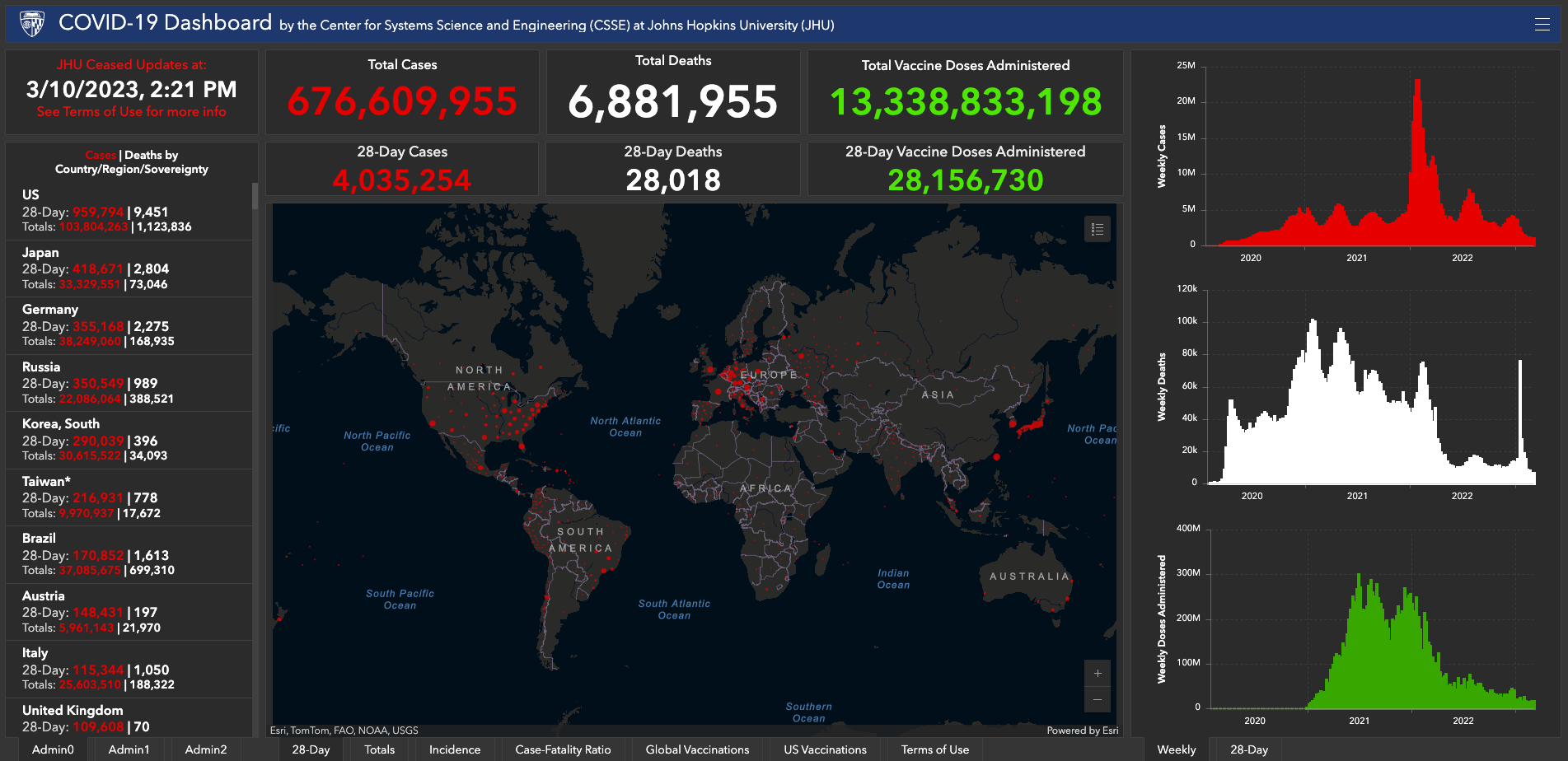} 
        \caption{JHU Dashboard.}
        \label{fig:jhu}
    \end{subfigure}
    \begin{subfigure}{0.5\textwidth}
        \includegraphics[width=0.9\linewidth]{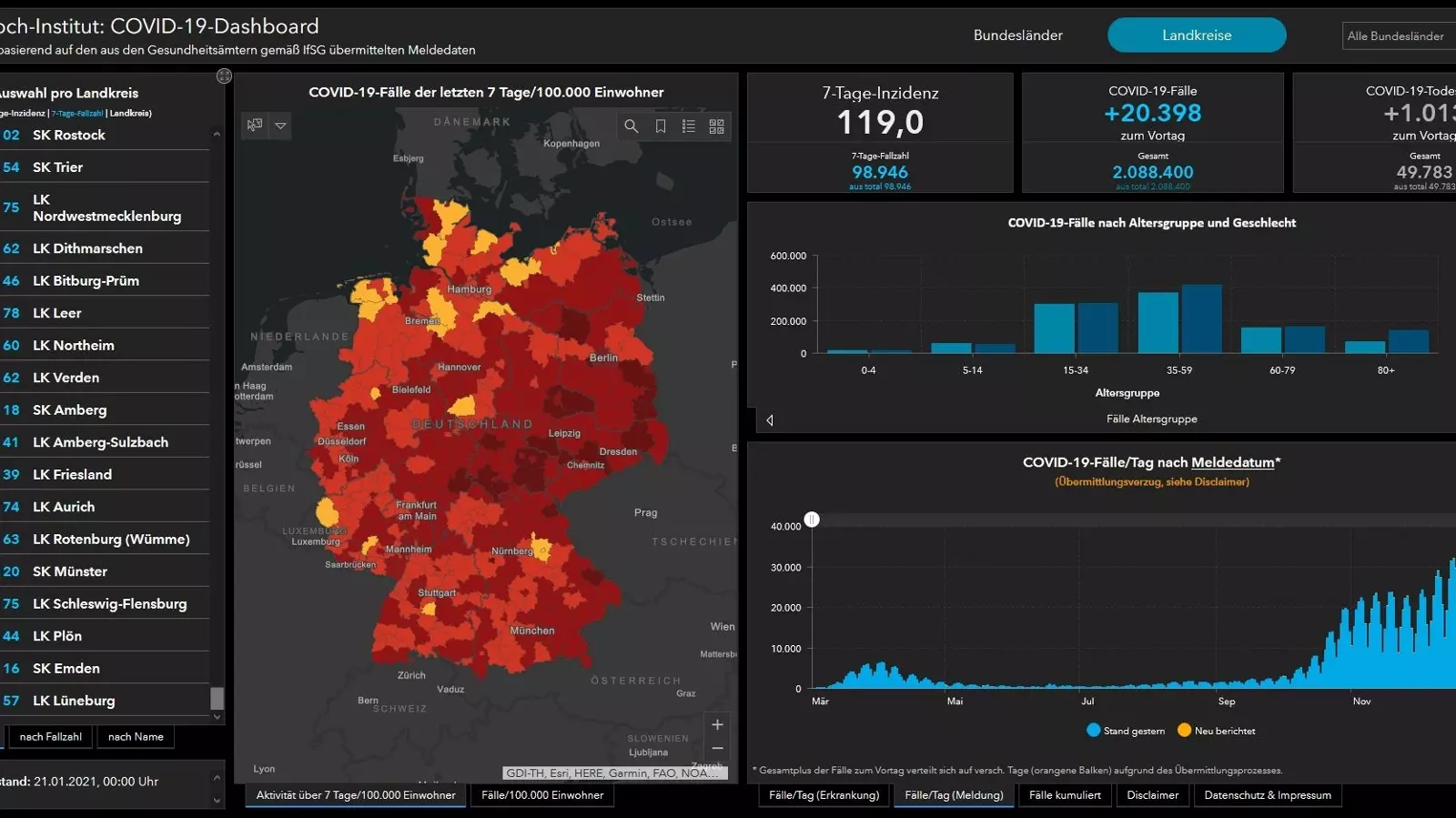} 
        \caption{RKI Dashboard.}
        \label{fig:rki}
    \end{subfigure}\\
        \begin{subfigure}{0.5\textwidth}
        \includegraphics[width=0.9\linewidth]{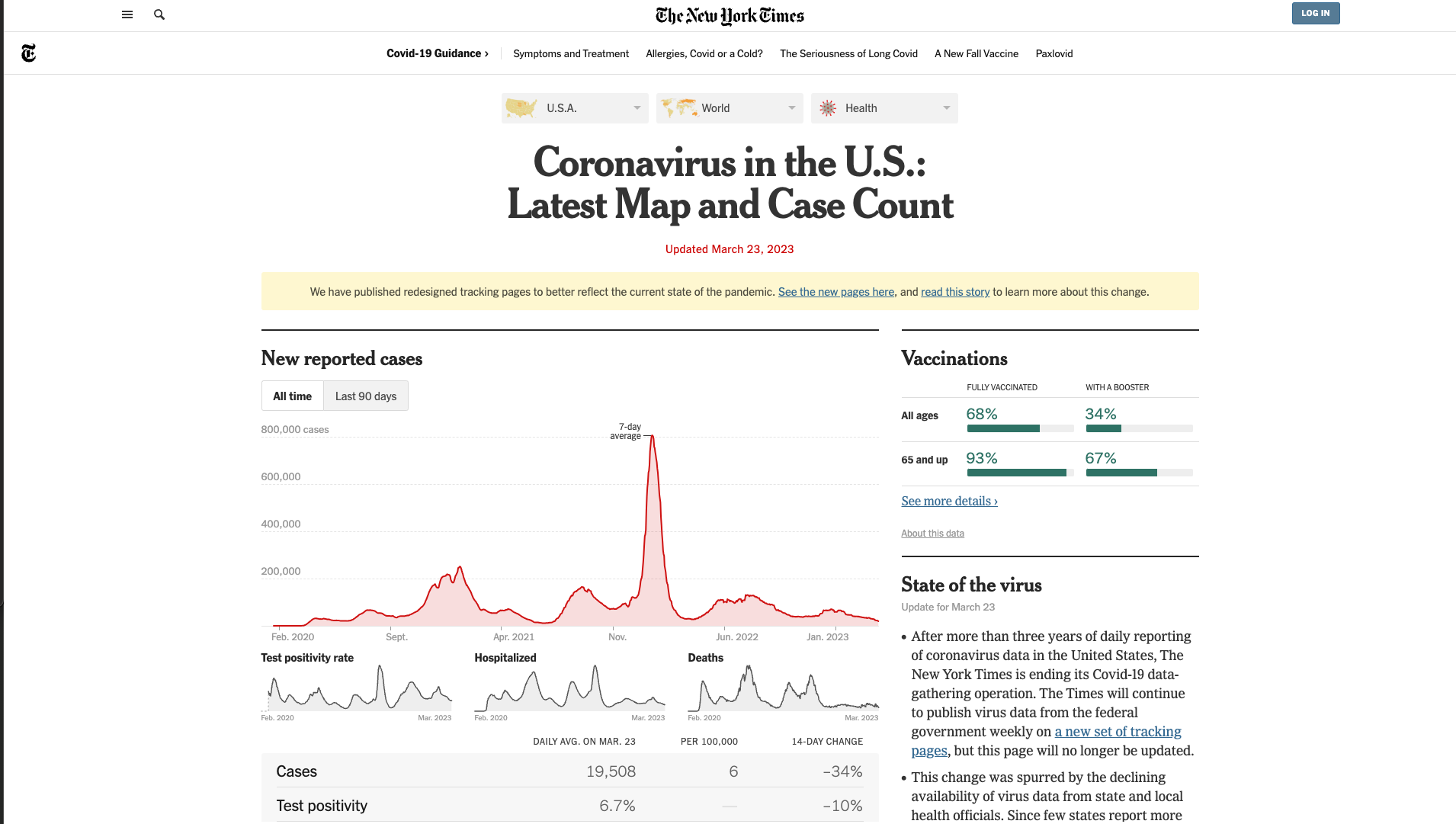} 
        \caption{NYT dashboard.}
        \label{fig:nyt}
    \end{subfigure}
    \begin{subfigure}{0.5\textwidth}
        \includegraphics[width=0.9\linewidth]{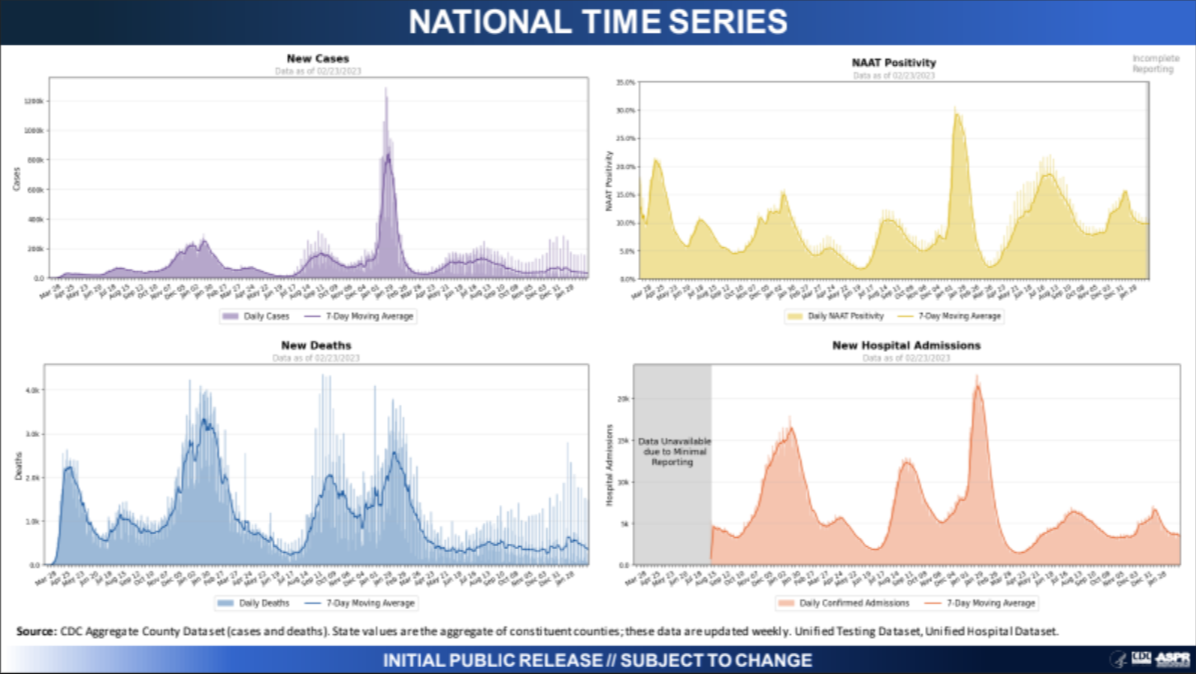} 
        \caption{White House Community Profile Report.}
        \label{fig:whitehouse}
    \end{subfigure}
    \caption{Four notable examples of the most popular dashboards seen during the Covid-19 pandemic. The difference between the designs of these dashboards is an effective indication of their intended scope and audience.}
\end{figure*}

The \textbf{John Hopkins University Center for Systems Science and Engineering} (JHU, see Figure~\ref{fig:jhu}) COVID-19 Dashboard was launched in January 2020 as one of the very first dashboards showing the pandemic data worldwide. The dashboard was used by a variety of user groups, ranging from politicians to media and the general public. It was built on the ESRI ArcGIS\footnote{\url{https://www.esri.com/en-us/arcgis/about-arcgis/overview}} software. The JHU Dashboard used custom-built web and data scraping techniques with manual data collection to integrate data from publicly accessible sources~\cite{dong2022johns}. A drawback of this solution was, especially at the beginning of the pandemic, the uncertainty of the data sources: for Germany, for instance, data from online services such as Worldometer\footnote{\url{https://www.worldometers.info/coronavirus/}} were used. These services themselves grabbed data from different websites, e.g., local newspapers. 

The German \textbf{Robert Koch-Institute} (RKI, see Figure~\ref{fig:rki}) dashboard reported the Covid-19 case data on the federal state and district levels. As with the JHU dashboard, it was technologically built on the ESRI ArcGIS software and was also used by a variety of user groups such as politicians, media and the general public. The major difference to the JHU dashboard is the data source. Only data of the official German reporting system were used in the RKI dashboard. 

\begin{figure}
    \centering
    \includegraphics[width=0.85\linewidth]{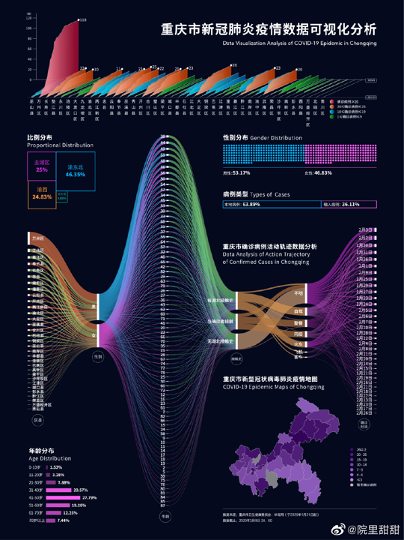}
    \caption[Caption for LOF]{Not all public dashboards used maps as the main visualization, like this example from Chongqing Municipal Health Commission\footnote{}.}
    \label{fig:covid_nomap}
\end{figure}

Apart from visualization developed by academia (JHU) and public authorities (RKI), media outlets also developed their own dashboards. One notable example is the \textbf{New York Times} (NYT, see Figure~\ref{fig:nyt}) dashboard. It showed data from state and local health agencies and the United States Department of Health and Human Services, aimed at informing the general public. 
Different from the first two examples, the NYT dashboard is a scrollable web page: graphs/maps/tables and explanatory text are shown in a sequential order by scrolling down the web page. The focus lies on inventing and experimenting with new visualizations and narrating the pandemic evolution in the form of journalistic storytelling. It is worth remarking that the majority of dashboards used an intuitive map visualization to show the high-level contagion evolution, but it was not an universal decision (see Figure~\ref{fig:covid_nomap}).

For completeness, we also mention \textbf{reports}, both confidential and for the general public, as forms of non-interactive dashboards.
Examples in this regard come from the public health agency of the German federal state of Lower Saxony (NLGA) and the COVID-19 Community Profile Report\footnote{\url{https://healthdata.gov/Health/COVID-19-Community-Profile-Report/gqxm-d9w9/about_data}} (see Figure~\ref{fig:whitehouse}) by the Data Strategy and Execution Group of the White House Covid-19 team. The former was used to inform the health ministry on a daily basis and was taken into consideration for the political actions to be taken. The second is a public report that mostly provided insight into the data trends throughout the COVID-19 pandemic to the general public.


To conclude, it is important to remark that for tracking the evolution of a pandemic there is no known ``silver bullet": different countries approach the problem also considering their own political and social context. 



\section{Interviewing Experts\\ Participants and Protocol}
Taking advantage of the unique context of a Dagstuhl seminar, we interviewed five experts from different backgrounds. The aims were to: (a) understand and document how dashboards had been essential to the work of health professionals and epidemiologists during the pandemic, and (b) identify takeaways for future dashboard design.

\textbf{Participants.} We conducted individual interviews with 5 experts (2 women and 3 men) and  refer to them as \textbf{i1-5} in the rest of the paper. Four of the interviewees worked in the German health sector in a media-facing role \textbf{(i1)}, a public health role for a state health authority \textbf{(i2)}, as a clinician \textbf{(i3)} and as a state epidemiologist \textbf{(i4)}.
The fifth \textbf{(i5)} was an epidemiologist who contributed to pandemic modeling in Scotland as a volunteer. 



\textbf{Protocol.} The interviews adapted a semi-structured approach from a previous study of data scientists~\cite{ruddle2023tasks}, and used the following main questions:
\begin{itemize}
\item What was your job/role during the pandemic?
\item Which dashboard(s) did you use?
\item What was the dashboard designed to let users do? Who were the users?
\item Data: What data was involved? Where did it come from?
\item Who developed the dashboard?
\item What issues, bottlenecks, pain points or challenges occurred?
\item Let yourself dream. What might have solved (or reduce) any of those difficulties, and what would the benefit have been to you?
\item How long did it take to learn how to use the dashboard? Was it worth the effort?
\item Do you still use dashboards in your current work? What for?
\end{itemize}
Each interview took 14 to 23 minutes.
The interviews were audio-recorded using Microsoft Teams, and the automatic transcription was cleaned up manually.
The interviewees gave verbal consent for the interviews to be recorded, transcribed and reported in research outputs.







\section{Interviewing Experts\\Results and discussion}

The interviewees primarily used third party, publicly available dashboards such as those provided by Johns Hopkins (see Figure~\ref{fig:jhu}), the RKI (federal-level data; see Figure~\ref{fig:rki}) and a national newspaper (Zeit Online). The clinician \textbf{(i3)} also used an internal dashboard showing hospital resources, and visualization techniques for transmission events from a previous research collaboration~\cite{baumgartl2020search}. The Scotland modeler was collaborating with RAMPVIS\footnote{\url{https://sites.google.com/view/rampvis/}}.

Most of the dashboards showed data aggregated at the level of countries or regions, but interviewees also typically either had access to some finer-grained data that was not in the public domain or worked with people who brought knowledge of such data to the collaboration.

\textbf{General Experience.} The interviewees indicated that it was straightforward to learn how to use the dashboards. Any issues were more likely to be about the data, though one interviewee shared that they where in the lucky situation that "\textit{I have a very multidisciplinary team}" so there "\textit{was always a person I can ask}" \textbf{(i3)}. 
The research dashboards were developed iteratively because "\textit{it took some time to get to a really good visualization that people can get the most information out of}" \textbf{(i3)}.

\textbf{Stakeholders and insights.}
The interviewees primarily used dashboards to: (1) answer specific questions from the media (radio, TV and newspapers), (2) assist policy makers, and (3) inform decision-making in a hospital. A typical media question was ``\textit{what does this mean for our population here?}" \textbf{(i4)}, but some questions were about the data itself. For example, "one day we had minus deaths" which triggered "phone calls the whole day" and comments such as "are [people] becoming alive again or what?" \textbf{(i2)}. The issue was caused by a country changing the criteria used to define a COVID-19 death, but a knock-on consequence was having to devote extra time to investigate such questions, to the detriment of people's other work. Similar issues occurred in previous pandemics (e.g., swine flu) but, in those days data was provided more slowly so it was possible to adjust for criteria changes over a few days by smoothing the numbers that were publicly released.

The first policy use was getting an impression from dashboard maps of how COVID-19 was spreading in other countries and regions, so ``you can calculate when it's coming" to your region and ``learn from their experiences" \textbf{(i3)}. Dashboards were a powerful aid for understanding how the pandemic was changing over time (``e\textit{pidemics run in waves, you can see that on the dashboard}" \textbf{(i4)}) and predictions about how interventions would prevent infections \textbf{(i5)}. When considering school closures, dashboards gave policy makers insights about questions such as ``\textit{how important are children in all of this? What} [...] \textit{measures to be taken and how can they be justified or not? What can the science say on this?}" \textbf{(i4)}. Interactive dashboards also enabled spatio-temporal differences to be investigated (e.g., between two districts of Bremen) and insights to be gained about infection rates in disadvantaged socio-economic areas \textbf{(i4)}. However, it was unnecessarily difficult to compare timelines of different regions - that is an area where dashboard functionality could have been improved \textbf{(i1)}.

Hospitals benefited from dashboards showing numbers of patients and resources (intensive care beds, ventilators, etc.). Not only did that help with the management of a hospital's own patients, it also led to them using spare capacity to treat patients from other regions and even some other countries \textbf{(i3)}.

\textbf{Issues and pain points.}
Inevitably, there were a variety of issues with the data that the interviewees had to use. Missing values are ``\textit{always an issue}" because those records may have a different distribution to the rest of the data, which makes imputation problematic \textbf{(i2)}. Sometimes entire variables were missing (e.g., the number of people tested), so incident rate charts were misleading \textbf{(i3)}. Where dashboards combined data from multiple sources then the data might be out of sync (e.g., delays providing mortality data) \textbf{(i4)}. Data provenance was a general issue - sometimes it was stated in the small print but other times not, leading to the dashboard providers being asked rather fundamental questions (``\textit{how actually, did you get to this number?}" \textbf{(i4)}).

\textbf{Dreams:}
One particular wish was to have access to finer-grained spatial data. In one case this was to ``\textit{have one layer further down for me to aggregate things in a way that I would like} [...] \textit{individual cities}" \textbf{i4} and in the other modeling was impeded because ``\textit{we also couldn't get the contact tracing data} [...] \textit{that local hotspots were kind of recording}" \textbf{(i5)}.

Some current software provides natural language interfaces or chart recommendation engines to help users create suitable visualizations. One interviewee dreamed that it ``\textit{would be really nice if I could type} [...] \textit{I now need the number of beds in intensive care units free}" and be automatically provided with a chart or ``\textit{perhaps a small map showing the regions and the data of the regions}", with the dashboard interpreting the query, finding suitable data and creating the visualizations \textbf{(i1)}.

\textbf{Do you still use dashboards?}
All of the interviewees still use dashboards. The most common use is for COVID-19 and other notifiable diseases that are slower-changing (e.g., ticks, tuberculosis and West Nile virus), with one interviewee commenting that the ``\textit{dashboards have improved}" because they now include the data pipeline and are more reliable \textbf{(i2)}. There is also a plan to adopt dashboards for other infectious agents, and especially carbapenem-producing antibacterials that are a World Health Authority priority \textbf{(i3)}.

\footnotetext{\url{https://www.jiqizhixin.com/articles/2020-03-23-4116}}

\textbf{Conclusive Remarks.} COVID-19 has led to new uses of dashboards and is changing working practices. One interviewee has already created her own animal infection dashboards, incorporating knowledge she gained from collaborating with the RAMPVIS visualization experts and, in the interview itself, suddenly realized that the infection tree visualizations could also be applied to animal routes \textbf{(i5)}. Epidemiologists now have greater appreciation of the benefits and are making conscious efforts use visualization more routinely and ``\textit{to visualize things in a better way}", as well as appreciating the benefits of making dashboards ``\textit{visually attractive for people to interact with}" \textbf{(i4)}.

\section{Takeaways}

Several recurrent themes emerged during the interviews, which we reflected upon and rephrased in the form of recommendations for future dashboard design. The recommendations are broadly applicable, and certainly not unique to COVID-19 dashboards.

The themes are not presented in a particular order, and we believe all to be equally important. However, weighing each of these features depends on the individual task at hand, and would therefore be the designer's duty to decide on the criteria to prioritize. Therefore, the following should be considered as a ``rulebook" of the desired properties of an effective dashboard, a set of principles meant to improve the design and functionality of future public health information dashboards by addressing key areas identified from the analysis of existing dashboards used during the COVID-19 pandemic.

\textit{Accessible Data Provenance}.
Clearly document data sources and methodologies used in dashboards. As we have seen with the JHU dashboard, an accurate data source is mandatory  to build trust between the system and the user. Other provenance issues ranged from the reasons data were missing, to an unambiguous description of the data source is advisable to enable users to assess the reliability of information themselves, and clarity about which criteria were used (e.g., to define COVID-19 deaths).

\textit{Integration of Data Sources}: Combining data from multiple sources is crucial to provide a comprehensive view, especially in crisis situations. 
However, inconsistency between sources adds to the challenge of assessing provenance.

\textit{Enhance Data Granularity}.
Depending on the context, providing a multiple level-of-detail access to data allows users to perform specific and localized analyses.

\textit{Data Sharing}. Three well-known ethics challenges in health are being: (a) granted permission to access data, (b) provided with data that has sufficient granularity for a given piece of analysis, and (c) allowed to link different datasets even after access has been obtained.
The authors have personal experience of the months and years of delays that those challenges sometimes cause to research, including preventing meaningful progress in COVID-19 projects during the pandemic.
At the heart of the challenges is a tendency for health organizations to be risk-averse, especially when information could single out specific individuals or groups of people.
The challenges directly conflict with some of the above themes by inhibiting the integration of different data sources and the provision of data at the granularity that is needed for specific analysis.
Our recommendations are twofold.
First, to establish principles for access to multi-source, fine-grained data and exemplars of its usage and benefits during normal (non-pandemic) times.
Second, to incorporate the inability to perform certain research into risk analysis for data ethics, for ethics boards to consider alongside security risks.

\textit{Consider the learning curve}. The selection of the idioms used in the dashboard should be accessible to a varied audience: therefore, the time required to ``learn" the individual views in a dashboard and how they interact with each other should be considered during the design process. For professional users, training sessions could be offered as well as detailed guides on how to use the system effectively.

\textit{Customizable by Design.}
Develop features that allow users to customize and manipulate data layers according to their needs. From our interviews, it turned out that a customizable dashboard was a common wish. These mostly concerned the data presentation, like  multiple scales and layers. We are aware that enabling customizability is a complex feature to enable in a visualization, however from our observations we argue that it could create a positive effect on the engagement of the users. Nonetheless, the extent of the customization is for the designers to decide and is closely related to the domain and context.

\textit{Critique by Experts}. Expert interviews are a valuable tool for creating visualizations both in the early design stage, as a preliminary step to identify and characterize the target domain, and when assessing the efficacy of the final product. In this context, based on our observations during the interviews, we found it crucial to incorporate feedback from both visualization experts (on aspects such as evaluation heuristics, color usage, and map region size) \textit{and} from domain experts from the target audience - possibly with varying levels of visualization literacy. As we have all experienced during the pandemic, ambiguous data representations led to fake news and misinformation: careful design and knowledge of the target audience expectations and characteristics might help mitigate this phenomenon in the future.

\textit{Human-machine interaction. } Guidance is a concept in visual analytics where visual cues support user queries and leverage human machine collaboration. It aims to support users in accomplishing their analytical goals and generating insights. In the context of pandemics, researchers deal with large, complex data - often noisy and with unclear provenance. Guidance, in this context, could highlight missing data, outliers, or relevant trends depending on the user goals, interactively adapting to the queries.

\textit{Focus on Interaction.}
Implement interactive elements to help users easily find and interpret the data they need. Interactivity was found to be positively received by all of the interviewed experts, also if they did not have a specific background in using visualization. Interaction design, generally, is a crucial element in dashboards.

\section{Conclusions}

In this paper, we had the rare opportunity to directly interface with experts and researchers of other fields than visualization who faced the challenges of understanding, communicating, and contrasting the pandemic using dashboards as part of their tools-of-trade. In this paper we gathered some of their thoughts, and organized them into a set of features we recommend considering when developing a dashboard in the context of pandemic visualization.

There are many ways in which this work, designed to be a first support for dashboard designers to establish their priorities in the design process, can be expanded and built upon. A further user study, with a larger user base could be used to confirm, expand, and question our findings. This paper also highlights the need to keep the momentum of the collaboration between visualization experts and public health communication, to avoid fake news and support responsible and effective decision making in tight and complex situations.


\section{ACKNOWLEDGMENTS}

We thank the other participants of Dagstuhl Seminar 24091, and especially the five people who took part in the interviews. 

\bibliographystyle{IEEEtran}
\bibliography{bibliography}

\begin{thebibliography}{10}
\providecommand{\url}[1]{#1}
\csname url@samestyle\endcsname
\providecommand{\newblock}{\relax}
\providecommand{\bibinfo}[2]{#2}
\providecommand{\BIBentrySTDinterwordspacing}{\spaceskip=0pt\relax}
\providecommand{\BIBentryALTinterwordstretchfactor}{4}
\providecommand{\BIBentryALTinterwordspacing}{\spaceskip=\fontdimen2\font plus
\BIBentryALTinterwordstretchfactor\fontdimen3\font minus \fontdimen4\font\relax}
\providecommand{\BIBforeignlanguage}[2]{{%
\expandafter\ifx\csname l@#1\endcsname\relax
\typeout{** WARNING: IEEEtran.bst: No hyphenation pattern has been}%
\typeout{** loaded for the language `#1'. Using the pattern for}%
\typeout{** the default language instead.}%
\else
\language=\csname l@#1\endcsname
\fi
#2}}
\providecommand{\BIBdecl}{\relax}
\BIBdecl

\bibitem{few2006information}
S.~Few, \emph{Information dashboard design: The effective visual communication of data}.\hskip 1em plus 0.5em minus 0.4em\relax O'Reilly Media, Inc., 2006.

\bibitem{kitchin2015knowing}
R.~Kitchin, T.~P. Lauriault, and G.~McArdle, ``Knowing and governing cities through urban indicators, city benchmarking and real-time dashboards,'' \emph{Regional Studies, Regional Science}, vol.~2, no.~1, pp. 6--28, 2015.

\bibitem{bach2023dashboard}
B.~Bach, E.~Freeman, A.~Abdul-Rahman, C.~Turkay, S.~Khan, Y.~Fan, and M.~Chen, ``Dashboard design patterns,'' \emph{IEEE Transactions on Visualization and Computer Graphics}, vol.~29, no.~1, pp. 342--352, 2023.

\bibitem{zhang2023dashboards}
Y.~Zhang, Y.~Sun, J.~D. Gaggiano, N.~Kumar, C.~Andris, and A.~G. Parker, ``Visualization design practices in a crisis: Behind the scenes with covid-19 dashboard creators,'' \emph{IEEE Transactions on Visualization and Computer Graphics}, vol.~29, no.~1, pp. 1037--1047, 2023.

\bibitem{sarikaya2018we}
A.~Sarikaya, M.~Correll, L.~Bartram, M.~Tory, and D.~Fisher, ``What do we talk about when we talk about dashboards?'' \emph{IEEE Transactions on Visualization and Computer Graphics}, vol.~25, no.~1, pp. 682--692, 2018.

\bibitem{wexler2017big}
S.~Wexler, J.~Shaffer, and A.~Cotgreave, \emph{The big book of dashboards: Visualizing your data using real-world business scenarios}.\hskip 1em plus 0.5em minus 0.4em\relax John Wiley \& Sons, 2017.

\bibitem{miksch2014triangle}
S.~Miksch and W.~Aigner, ``A matter of time: Applying a data–users–tasks design triangle to visual analytics of time-oriented data,'' \emph{Computers \& Graphics}, vol.~38, pp. 286--290, 2014.

\bibitem{dong2022johns}
E.~Dong, J.~Ratcliff, T.~D. Goyea, A.~Katz, R.~Lau, T.~K. Ng, B.~Garcia, E.~Bolt, S.~Prata, D.~Zhang \emph{et~al.}, ``The {J}ohns {H}opkins {U}niversity {C}enter for {S}ystems {S}cience and {E}ngineering {COVID-19 D}ashboard: Data collection process, challenges faced, and lessons learned,'' \emph{The Lancet Infectious Diseases}, vol.~22, no.~12, pp. e370--e376, 2022.

\bibitem{ruddle2023tasks}
R.~A. Ruddle, J.~Cheshire, and S.~J. Fernstad, ``Tasks and visualizations used for data profiling: A survey and interview study,'' \emph{IEEE Transactions on Visualization and Computer Graphics}, 2023.

\bibitem{baumgartl2020search}
T.~Baumgartl, M.~Petzold, M.~Wunderlich, M.~Hohn, D.~Archambault, M.~Lieser, A.~Dalpke, S.~Scheithauer, M.~Marschollek, V.~Eichel \emph{et~al.}, ``In search of patient zero: Visual analytics of pathogen transmission pathways in hospitals,'' \emph{IEEE Transactions on Visualization and Computer Graphics}, vol.~27, no.~2, pp. 711--721, 2020.

\end{thebibliography}


\begin{IEEEbiography}{A. Arleo}{\,} is an Assistant Professor at the Visualization Cluster at the Eindhoven University of Technology. He is currently leading a research project about visualization of diffusion processes on temporal contact tracing networks at the Vienna University of Technology. Contact him at a.arleo@tue.nl. 
\end{IEEEbiography}

\begin{IEEEbiography}{R. Borgo}{\,} is Reader in Data Visualization at the Informatics Department at King’s College London (KCL). She is  Head of the Human Centred Computing research group. Her research focus is on Human Factors in Visualization. Contact her at rita.borgo@kcl.ac.uk.
\end{IEEEbiography}

\begin{IEEEbiography}{J. Kohlhammer}{\,} is with Fraunhofer IGD and TU Darmstadt, Germany. He leads an applied research department on visual analytics with applications in health, engineering and cyber-security. Contact him at joern.kohlhammer@igd.fraunhofer.de.
\end{IEEEbiography}

\begin{IEEEbiography}{R. Ruddle} {\,} is a Professor of Computing at the University of Leeds, UK. He co-led the Leeds analysis team in the Alan Turing Institute DECOVID project. Contact him at r.a.ruddle@leeds.ac.uk.
\end{IEEEbiography}

\begin{IEEEbiography}{H. Scharlach}{\,} is head of communications at the Public Health Agency of Lower Saxony, Germany and spokesperson of the working group on Health Geography in the German Society of Geography. Contact him at holger.scharlach@nlga.niedersachsen.de.
\end{IEEEbiography}

\begin{IEEEbiography}{X. Yuan}{\,} is a tenured faculty member in the School of Electronics Engineering and Computer Science. His primary research interests are in the field of scientific visualization, information visualization and visual analytics. Contact him at xiaoru.yuan@pku.edu.cn.
\end{IEEEbiography}

\end{document}